\documentclass[twocolumn]{jpsj2} 
\title{
On the Meissner Effect of the Odd-Frequency Superconductivity with Critical Spin Fluctuations:\\
Possibility of Zero Field FFLO pairing}

\author{Yuki \textsc{Fuseya}
\thanks{E-mail: fuseya@mp.es.osaka-u.ac.jp}
and Kazumasa \textsc{Miyake}
}

\inst{Department of Materials Engineering Science, 
Osaka University, 1-3 Machikaneyama, Toyonaka, Osaka 560-8531 
}

\abst{
	We investigate the influence of critical spin fluctuations on electromagnetic responses in the odd-frequency superconductivity.
	It is shown that the Meissner kernel of the odd-frequency superconductivity is strongly reduced by the critical spin fluctuation or the massless spin wave mode in the antiferromagnetic phase.
	These results imply that the superfluid density is reduced, and the London penetration depth is lengthened for the odd-frequency pairing.
	It is also shown that the zero field Flude-Ferrell-Larkin-Ovchinnikov pairing is spontaneously realized both for even- and odd-frequency in the case of sufficiently strong coupling with low lying spin-modes.
}

\kword{odd-frequency superconductivity, Meissner effect, spin fluctuation, superfluid density, penetration depth, zero-field FFLO}

\usepackage{bm}
\usepackage{mathrsfs}

\newcommand{\br}{\bm r}
\newcommand{\bp}{\bm p}
\newcommand{\bq}{\bm q}

\newcommand{\im}{{\rm i}}

\newcommand{\mG}{\mathscr{G}}
\newcommand{\mF}{\mathscr{F}}
\newcommand{\mD}{\mathscr{D}}
\newcommand{\up}{\uparrow}
\newcommand{\down}{\downarrow}
\newcommand{\ve}{\varepsilon}
\newcommand{\NF}{N_{\rm F}}
\newcommand{\eF}{\ve_{\rm F}}
\newcommand{\Tc}{T_{\rm c}}
\newcommand{\oddw}{odd-$\omega$ }
\newcommand{\evenw}{even-$\omega$ }

\begin{document}
\maketitle

\section{Introduction} 
	%
	%
	%
	%
	%
	Odd-frequency superconductivity, whose gap function is odd in frequency, is an old but new class of superconductivity.
	It was introduced originally by Berezinskii for triplet pairing\cite{Berezinskii1974}, and by Balatsky and Abrahams for singlet pairing\cite{Balatsky1992}.
	Soon after the Balatsky-Abrahams, the possibility of odd-frequency superconductivity (odd-$\omega$ SC) was discussed  in a wide variety of models, e.g.,  the Kondo lattice model,\cite{Emery1992,Coleman1993,Coleman1994,Coleman1995,Coleman1997} the square-\cite{Bulut1993} and triangular lattice\cite{Vojta1999} Hubbard model, and the $t$-$J$ model.\cite{Balatsky1993}

	The experimental evidence on the \oddw SC was first pointed out by the authors.\cite{Fuseya2003}
	In Ce-based heavy fermion superconductors, such as CeCu$_2$Si$_2$ and CeRhIn$_5$, several experiments on the nuclear magnetic relaxation rate\cite{Ishida1999,YKawasaki2002,SKawasaki2003} and specific heat\cite{Modler1995,Fisher2002} exhibited behaviors of gapless SC near the antiferromagnetic (AF) quantum critical point (especially in the AF phase).
	The origin of this gapless SC had long been a mystery.
	In Ref. \citen{Fuseya2003}, the authors showed that the \oddw SC is realized due to spin-fluctuation near the AF quantum critical point and/or in the AF state by solving the Bethe-Salpeter equation\cite{Salpeter1951} of SC  based on the itinerant-localized duality theory\cite{Miyake1991,Kuramoto1992}.
	%
	The phase diagram so obtained agrees well with the experimental phase diagram, considering the fact that the \oddw SC exhibits the gapless behavior.
	Accordingly, the origin of the gapless SC in Ce-compounds can be understood as the \oddw SC.

	After this suggestion, detailed investigations have been made into the triangular-type lattice,\cite{Yada2008,Shigeta2009,Shigeta2010} and into the effective model of U-compounds\cite{Hotta2009}, both of which are revealed to possess the \oddw state under some realistic conditions.
	It now appears probable that the \oddw  SC can be realized in surprisingly familiar situations.
	The concept of the \oddw SC is so general that nowadays it has expanded into various fields, e.g., magnetic or density wave order\cite{Balatsky1995,Pivovarov2001}, superconducting junctions\cite{Buzdin2005,Bergeret2005,Bergeret2001,Eschrig2003,Tanaka2007a,Asano2007a,Tanaka2007b,Braude2007,Yokoyama2007,Asano2007b,Tanaka2007c,Eschrig2007,Asano2007c,Fominov2007,Eschrig2008,Tanaka2008,Linder2009,Volkov2010}, vortex core\cite{Yokoyama2008,Tanuma2009}, proximity effect superfluid $^3$He\cite{Higashitani2009}, and cold atoms\cite{Kalas2008}.

	Besides the studies on the possible existence of \oddw SC, the thermodynamic stability or the sign of the Meissner response had been still remaining unclear.
	Within a naive calculation, the Meissner kernel of the \oddw pairing seems to have an opposite sing from that of the usual (\evenw) pairing, suggesting the thermodynamically unstable superconducting states.
	Very recently, this puzzle was solved by the appropriate treatment of the gap function with retardation in the path-integral formalism\cite{Solenov2009,Kusunose2010}.
	According to these works, the \oddw SC is thermodynamically stable, and  exhibits the conventional Meissner effect.
	However, these arguments are restricted only to the coherent part (without the correlations due to the incoherent part of the quasiparticle).
	By contrast, previous studies\cite{Emery1992,Coleman1993,Coleman1994,Coleman1995,Coleman1997,Bulut1993,Vojta1999,Fuseya2003,Yada2008,Shigeta2009,Shigeta2010,Hotta2009} demonstrated that the \oddw solutions are realized only when the Cooper pairs are mediated by a certain strong spin fluctuation, strongly suggesting the relevant contribution of the incoherent part.
	So we have to discuss carefully the Meissner effect of the \oddw pairing considering the contributions from the incoherent part, which are neglected in the previous arguments\cite{Solenov2009,Kusunose2010}.
	In this paper, we consider the incoherent corrections due to critical spin fluctuations which are crucial ingredients for realizing the \oddw SC.

	In \S 2, the ordinary arguments for the electromagnetic response, where the incoherent part is neglected, are briefly reviewed.
	In \S 3, we introduce spin fluctuations in a general form and investigate the Meissner effect by calculating the current-current response function with the first order correction of the critical spin fluctuations or spin wave modes in the AF state.
	There, we examine the singlet pairing in the AF phase, where the realization of the \oddw is guaranteed by the previous work\cite{Fuseya2003}.
	The results indicate that the Meissner kernel is reduced due to the correction of such spin fluctuations.
	Next, in \S 4, we study the triplet pairing with ferromagnetic spin fluctuations in the same manner as in \S 3.
	It is shown that the Meissner effect is also reduced in triplet pairings.
	In \S 5, we give another argument of the electromagnetic response by calculating the superconducting density, which is given by the spatially gradient term in the free energy.
	Up to the first order of the critical spin fluctuations, its result is equal to that obtained by the current-current response function as expected.
	With this procedure, we carry out the calculation of higher order corrections up to the third order.
	Then, we conclude that the Meissner kernel is strongly reduced both for even- and \oddw pairing, even if we take into account the infinite order of corrections.
	Our results also suggest that the Flude-Ferrell-Larkin-Ovchinnikov (FFLO) pairing is spontaneously realized without a magnetic field in the case of sufficiently strong coupling.
	Finally, in \S 6, we discuss implications of the present results.

\section{Electromagnetic response of the coherent part}

\begin{figure}[tb]
\begin{center}
\includegraphics[width=7cm]{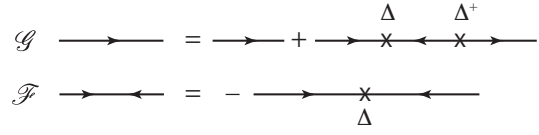}
\end{center}
\caption{Diagrammatic expressions of the expansion of the normal $\mG$ and anomalous Green function $\mF$.}
\label{GFexpand}
\end{figure}

	%
	In this section, we reintroduce the Meissner kernel in the Ginzburg-Landau (GL) region with the coherent part only in a convenient form for later discussions.
	The relation between the current density and the vector potential has the following form\cite{AGD}
\begin{align}
	{\bm j}(q) &= -Q(q) \bm{A}(q) \nonumber\\
	&=
	-\frac{Ne^2}{m}{\bm A}(q)
	-\frac{2e^2T}{m^2}
	\sum_{\bp, n} \bp \left[ \bp \cdot{\bm A}(q) \right]
	\nonumber\\&\times
	\left[
	\mG (p_+) \mG (p_-) + \mF(p_+) \mF^+ (p_-)
	\right],
\end{align}
	where $p\equiv (\bp, \ve_n)$, $p_\pm\equiv p\pm q/2$, $\mG$ and $\mF^{(+)}$ are the normal Green function and the anomalous Green function, respectively.
	$N$ is the number of quasiparticles with a charge $e$ and a mass $m$.
	The kernel $Q_{xx}(q)$ is given basically by the current-current response function, which consists of a particle-hole diagram.
	(Note that the spatially gradient term in the free energy consists of particle-particle diagrams and will be discussed in \S 5.) 
	%
	%
	%
	%
	These Green functions are given as
\begin{align}
	\mG (\bp, \ve_n ) &= -\frac{\im \ve_n + \xi_{\bp}}{\ve_n^2 + \xi_{\bp}^2 + |\Delta_{\mu} (\bp, \ve_n)|^2},\\
	\mF^+ (\bp, \ve_n ) &= \frac{\Delta^+_{\mu} (\bp, \ve_n )}
	{\ve_n^2 + \xi_{\bp}^2 + |\Delta_{\mu} (\bp, \ve_n)|^2},
\end{align}
	both for the \evenw ($\Delta_{\rm e}$) and \oddw ($\Delta_{\rm o}$)\cite{Kusunose2010}.
	In the GL region, i.e., near $\Tc$, we can expand them with respect to $\Delta_{\mu}$.
	Up to the second order in $\Delta_{\mu}$, we have
\begin{align}
	\mG(p)&\simeq G(p) + 
	\frac{\im \ve_n + \xi_{\bp}}{(\ve_n^2 + \xi_{\bp}^2)^2}|\Delta_{\mu} (p)|^2
	\nonumber\\
	&= G(p) + G(-p)\left[ G(p)\right]^2 |\Delta_{\mu}(p)|^2,
	\label{2ndG}\\
	\mF(p)&\simeq
	-\frac{\Delta^{(+)}_{\mu} (p)}{\ve_n^2 + \xi_{\bp}^2}
	=-G(p)G(-p) \Delta_{\mu}^{(+)}(p),
	\label{2ndF}
\end{align}
	where $G$ is the Green function in the normal state $G(p)=\left[ \im \ve_n -\xi_{\bp}\right]^{-1}$.
	The diagrammatic expressions of these expansions are shown in Fig. \ref{GFexpand}. 
	%
	The Meissner effect can be demonstrated in the limit of $q\to0$.
After some straightforward manipulations, we have
\begin{align}
Q_{xx}(0)
&=
2e^2 T\sum_{\bp, n}
\left( \frac{p_x}{m} \right)^2
\nonumber\\ &\times
\left\{
-2 [G (p)]^3 G (-p)
+ [G(p)G(-p)]^2
\right\} \left|\Delta_{\mu}(\ve_n)\right|^2
\nonumber\\
&=\frac{Ne^2}{m} \pi T \sum_{n} \frac{\left|\Delta_{\mu}(\ve_n)\right|^2}{|\ve_n|^3}>0,
\label{7}
\end{align}
which leads to the ordinary (diamagnetic) Meissner effect\cite{AGD}.
(Here, we have used the relation $N=p_{F}^{3}/3\pi^{2}$.)
This expression, however, does not include the effect of the incoherent part, or the spin fluctuations, which is indispensable for the realization of the \oddw SC.
Only from eq. (\ref{7}), we cannot see how the Meissner effect and electromagnetic responses will be modified by the indispensable corrections of the incoherent part.

It should be noted here that there is no difference between the even- and \oddw pairing in the calculation of the Meissner kernel (except for the $\ve_n$-dependence of $\Delta (\ve_n)$), since the structures of $\mG$ and $\mF$ are the same.
In other words, the classification of even- and \oddw is that with respect to the relative coordinate, $p$, whereas the electromagnetic response is to the center of mass coordinate, $q$, so that the theoretical framework of the electromagnetic response is basically the same for even- and \oddw pairings. 	
	%
	
	%

\section{Singlet pairings}
\subsection{Model}

	Our purpose here is to calculate the corrections of fluctuations to the Meissner effect.
	For this purpose, we start from a simple and essential model.
	Judging from the previous studies\cite{Emery1992,Coleman1993,Coleman1994,Coleman1995,Coleman1997,Vojta1999,Yada2008,Shigeta2009,Bulut1993,Balatsky1993,Fuseya2003}, the \oddw pairing tends to be mediated by spin fluctuations, which originate from the local component, i.e., the incoherent part.
	Therefore, we introduce the following low-energy effective action on the basis of the itinerant-localized duality theory\cite{Miyake1991,Kuramoto1992}:
\begin{align}
	S&=\sum_{\bp, \alpha} \int_0^{1/T} \! \! d\tau 
	\bar{\psi}_{\bp, \alpha} (\tau) (\partial_\tau + \xi_{\bp})\psi_{\bp, \alpha} (\tau)
	\nonumber\\
	&-g^2 \sum_{\bq} \int_0^{1/T} \! \! d\tau \int_0^{1/T} \! \! d\tau'
	\mD(\bq, \tau - \tau') {\bm S}(\bq, \tau)\cdot {\bm S}(-\bq, \tau'),
\end{align}
	where $\bar{\psi}_{\bp, \alpha}(\tau)\equiv e^{\tau \mathscr{H}}\psi_{\bp, \alpha}^\dagger e^{-\tau \mathscr{H}}$, and $g$ is a coupling constant.
	The spin density is given by
\begin{align}
	{\bm S}(\bq, \tau)=\frac{1}{2}\sum_{\bp, \alpha, \beta}
	\bar{\psi}_{\bp+\bq/2, \alpha}(\tau) {\bm \sigma}_{\alpha\beta}\psi_{\bp-\bq/2, \beta}(\tau),
\end{align}
	where ${\bm \sigma}$ denotes the Pauli matrix.
	%
	%
	%
	%
	%
	The actual \oddw pairing, i.e., the gapless superconductivity, has been reported mainly in the coexistence phase of superconductivity and antiferromagnetism\cite{YKawasaki2002,SKawasaki2003,Modler1995,Fisher2002}.
	For such a situation, the most dominant fluctuation is the transverse spin-fluctuation due to the spin wave as
\begin{align}
	\mD (\bq, \omega_m) = \frac{g^2 T_{\rm N}}{v_{\rm s}^2 \hat{\bq}^2 + |\omega_m|^2},
\end{align}
	where $g$ is the coupling constant, $T_{\rm N}$ expresses the energy scale of the AF transition temperature, $v_{\rm s}$ is the velocity of the spin-wave, and $\hat{\bq}=\bq-{\bm Q}$ (${\bm Q}$ is the magnetic order vector).
	This type of spin fluctuation has already been shown to realize the \oddw SC in the AF phase\cite{Fuseya2003}.
	When we consider in the doubly folded Brillouin zone, $\hat{\bq}$ is replaced by $\bq$, so that $\mD (\bq, \omega_m)$ is singular for $\bq\sim 0$ and $\omega_m \sim 0$.
	We use this transverse spin-fluctuation for singlet pairings.

	%
	%
	%
	%
	
\begin{figure}[tb]
\begin{center}
\includegraphics[width=8cm]{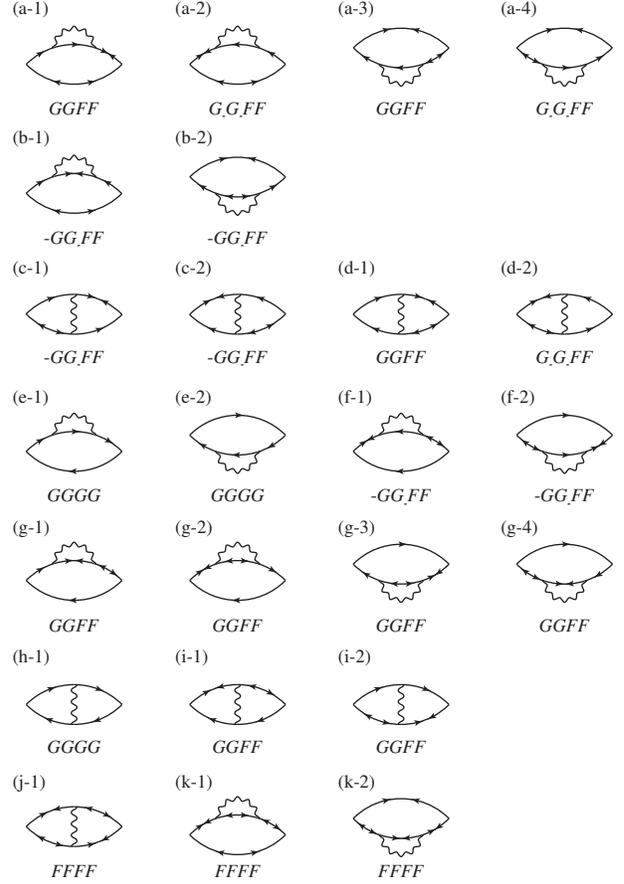}
\end{center}
\caption{Diagrammatic expressions of the current-current correlation with correction of fluctuations.}
\label{GFdiagrams}
\end{figure}

	The corrections of the spin-fluctuation to the kernel $Q_{xx}$ are
\begin{align}
	&\Delta Q_{xx}(q)=
	2e^{2}T\sum_{p, q'} \mD(q') |\Delta_\mu (p)|^2
	\nonumber\\
	&\times
	\bigl[
	-v_{p}v_{-p}
	\mG(p_{+})\mG(p_{+}-q')\mF_{\up \down}^{+}(p_{-})\mF_{\down \up}(p_{+})
	\tag{a-1}
	\nonumber\\
	&-v_{p}v_{-p}
	\mG(-p_{+})\mG(-p_{+}+q')\mF_{\up\down}^{+}(p_{-})\mF_{\down\up}(p_{+})
	\tag{a-2}
	\nonumber\\
	&-v_{p}v_{-p}
	\mG(p_{-})\mG(p_{-}+q')\mF_{\up\down}^{+}(p_{-})\mF_{\down\up}(p_{+})
	\tag{a-3}
	\nonumber\\
	&-v_{p}v_{-p}
	\mG(-p_{-})\mG(-p_{-}+q')\mF_{\up\down}^{+}(p_{-})\mF_{\down\up}(p_{+})
	\tag{a-4}
	\nonumber\\
	&+v_{p}v_{-p}
	\mG(p_{+})\mG(-p_{+})\mF_{\up \down}^{+}(p_{-})\mF_{\down \up}(-p_{+}+q')
	\tag{b-1}
	\nonumber\\
	&+v_{p}v_{-p}
	\mG(p_{-})\mG(-p_{-})\mF_{\up\down}^{+}(-p_{-}+q')\mF_{\down\up}(p_{+})
	\tag{b-2}
	\nonumber\\
	&+v_{p}v_{-p+q'}
	\mG(p_{+})\mG(-p_{-}+q')\mF_{\up\down}^{+}(p_{-})\mF_{\down\up}(-p_{+}+q')
	\tag{c-1}
	\nonumber\\
	&+v_{p}v_{-p+q'}
	\mG(p_{-})\mG(-p_{+}+q')\mF_{\up \down}^{+}(-p_{-}+q')\mF_{\down \up}(p_{+})
	\tag{c-2}
	\nonumber\\
	&-v_{p}v_{-p+q'}
	\mG(p_{+})\mG(p_{-})\mF^{+}_{\up\down}(-p_{+}+q')\mF_{\down \up}(-p_{-}+q')
	\tag{d-1}
	\nonumber\\
	&-v_{p}v_{-p+q'}
	\mG(-p_{+}+q')\mG(-p_{-}+q')\mF_{\up\down}^{+}(p_{-})\mF_{\down\up}(p_{+})
	\tag{d-2}
	\nonumber\\
	&+v_{p}v_{p}
	\mG(p_{+})\mG(p_{+})\mG(p_{-})\mG(p_{+}-q')
	\tag{e-1}
	\nonumber\\
	&+v_{p}v_{p}
	\mG(p_{+})\mG(p_{-})\mG(p_{-})\mG(p_{-}+q')
	\tag{e-2}
	\nonumber\\
	&-v_{p}v_{p}
	\mG(p_{-})\mG(-p_{+}+q')\mF_{\up\down}^{+}(p_{+})\mF_{\down\up}(p_{+})
	\tag{f-1}
	\nonumber\\
	&-v_{p}v_{p}
	\mG(p_{+})\mG(-p_{-}+q')\mF_{\up\down}^{+}(p_{-})\mF_{\down\up}(p_{-})
	\tag{f-2}
	\nonumber\\
	&+v_{p}v_{p}
	\mG(p_{+})\mG(p_{-})\mF_{\up \down}^{+}(p_{+})\mF_{\down \up}(-p_{+}+q')
	\tag{g-1}
	\nonumber\\
	&+v_{p}v_{p}
	\mG(p_{-})\mG(p_{+})\mF_{\up\down}^{+}(-p_{+}+q')\mF_{\down\up}(p_{+})
	\tag{g-2}
	\nonumber\\
	&+v_{p}v_{p}
	\mG(p_{+})\mG(p_{-})\mF_{\up\down}^{+}(-p_{-}+q')\mF_{\down\up}(p_{-})
	\tag{g-3}
	\nonumber\\
	&+v_{p}v_{p}
	\mG(p_{+})\mG(p_{+})\mF_{\up\down}^{+}(p_{-})\mF_{\down\up}(-p_{-}-q')
	\tag{g-4}
	\nonumber\\
	&+v_{p}v_{p-q'}
	\mG(p_{+})\mG(p_{-})\mG(p_{+}-q')\mG(p_{-}-q')
	\tag{h-1}
	\nonumber\\
	&+v_{p}v_{p-q'}
	\mG(p_{-})\mG(p_{+}-q')\mF_{\up \down}^{+}(-p_{+}+q')\mF_{\down \up}(p_{+})
	\tag{i-1}
	\nonumber\\
	&+v_{p}v_{p-q'}
	\mG(p_{+})\mG(p_{+}-q')\mF_{\up\down}^{+}(p_{-})\mF_{\down\up}(-p_{-}+q')
	\tag{i-2}
	\nonumber\\
	&+v_{p}v_{p-q'}
	\mF_{\up\down}^{+}(p_{-})\mF_{\down\up}(p_{+})\mF_{\up \down}^{+}(-p_{+}+q')\mF_{\down \up}(-p_{-}+q')
	\tag{j-1}
	\nonumber\\
	&-v_{p}v_{-p}
	\mF_{\up\down}^{+}(p_{-})\mF_{\down\up}(p_{+})\mF_{\up\down}^{+}(-p_{+}+q')\mF_{\down\up}(p_{+})
	\tag{k-1}
	\nonumber\\
	&-v_{p}v_{-p}
	\mF_{\up\down}^{+}(p_{-})\mF_{\down\up}(p_{+})\mF_{\up\down}^{+}(p_{-})\mF_{\down\up}(-p_{-}+q')
	\tag{k-2}
	\bigr],
	\nonumber
\end{align}
	where $v_p=p_x /m$ is the velocity of quasiparticles.
	The corresponding diagrams are shown in Fig. \ref{GFdiagrams}.
	%
	In the limit of $q\to0$, we have
\begin{align}
	&\Delta Q_{xx}(0)
	=2e^{2}T\sum_{p,q'} v_{p}^{2}\mD(0) |\Delta_\mu (p)|^2
	\nonumber\\
	&\times\biggl[
	3\mG(p)\mG(p)\mG(p)\mG(p)
	+3\mG(-p)\mG(-p)\mF_{\up \down}^{+}(p)\mF_{\down \up}(p)
	\nonumber\\&
	+2\mG(p)\mG(p)\mF_{\up \down}^{+}(p)\mF_{\down \up}(p)
	+\mG(p)\mG(p)\mF_{\up \down}^{+}(-p)\mF_{\down \up}(-p)
	\nonumber\\&
	+3\mG(p)\mG(p)\mF_{\up \down}^{+}(p)\mF_{\down \up}(-p)
	+3\mG(p)\mG(p)\mF_{\up \down}^{+}(-p)\mF_{\down \up}(p)
	\nonumber\\&
	-2\mG(p)\mG(-p)\mF_{\up \down}^{+}(p)\mF_{\down \up}(p)
	-2\mG(p)\mG(-p)\mF_{\up \down}^{+}(p)\mF_{\down \up}(-p)
	\nonumber\\&
	-2\mG(p)\mG(-p)\mF_{\up \down}^{+}(-p)\mF_{\down \up}(p)
	\nonumber\\&
	+\mD(q')\mF_{\up \down}^+(p)\mF_{\down \up}(p)\mD(q')\mF_{\up \down}^+(-p)\mF_{\down \up}(-p)
	\nonumber\\&
	+\mD(q')\mF_{\up \down}^+(p)\mF_{\down \up}(p)\mD(q')\mF_{\up \down}^+(-p)\mF_{\down \up}(p)
	\nonumber\\&
	+\mD(q')\mF_{\up \down}^+(p)\mF_{\down \up}(p)\mD(q')\mF_{\up \down}^+(p)\mF_{\down \up}(-p)
	\biggr].
\end{align}
	Here, we have put $q' \to 0$ since only $\mD(\bq, \omega_m)$ with $\omega_m=0$ is important in the Ginzburg-Landau region.
	%
	Up to the second order of $\Delta(p)$ (eqs. (\ref{2ndG})-(\ref{2ndF})),
\begin{align}
	\Delta Q_{xx}(0)
	&=
	2e^{2}T\sum_{p}v_{p}^{2}\mD_0 |\Delta_\mu (p)|^2
	\nonumber\\ &\times 
	\left[
	-12G_p^{5}G_{-p}^{1}
	+12G_p^{4}G_{-p}^{2}
	-6G_p^{3}G_{-p}^{3}
	\right]
	\label{13}\\
	&\simeq 
	2e^{2}T\sum_{n}
	\mD_0 |\Delta_\mu (\ve_{n})|^{2}
	\frac{2N_{F}\ve_{F}}{3m}
	\nonumber\\&\times
	\left\{
	-12\left(\frac{\pi}{16}\right)
	+12\left(-\frac{\pi}{4}\right)
	-6\left(\frac{3\pi}{8}\right)
	\right\} \frac{1}{|\ve_{n}|^{5}}
	\nonumber\\&
	=2e^{2}T\sum_{n}
	\mD_0 |\Delta_\mu(\ve_{n})|^{2}
	\frac{2N_{F}\ve_{F}}{3m}
	\left( -6\pi \right)
	\frac{1}{|\ve_{n}|^{5}}
	\nonumber\\&
	\simeq\frac{e^{2}N}{m}T\sum_{n}
	\mD_0 |\Delta_\mu (\ve_{n})|^{2}
	\left( -6\pi \right)
	\frac{1}{|\ve_{n}|^{5}},
	\label{15}
\end{align}
	where $N_{F}\equiv mp_{F}/2\pi^{2}$, $\ve_{F}\equiv p_{F}^{2}/2m$, and $G(p)$'s are abbreviated as $G_p$'s. 
	Note that $\Delta(\ve_n)$ includes the factor from the momentum integration of $\Delta (p)$ as $\Delta (\ve_n) \sim \sum_{\bp} \Delta(\bp, \ve_n) \Phi(\bp, \ve_n)$, where $\Phi (\bp, \ve_n)$ is some function, and  $\mD_0$ includes the factor from the frequency summation  ($T\sum_{q'}$).
	At around $\Tc$, the dominant contribution of $\Delta (\ve_n)$ comes only from $\ve_0 =\pi \Tc$, so that we can approximate as 
	\begin{align}
	\mD_0=T\sum_q g^2 T_{\rm N}/v_{\rm s}^2 q^2 \sim g^2 \Tc / T_{\rm N}.
	\end{align}
	(Here we use $v_{\rm s} q_{\rm c} \sim T_{\rm N}$, where $q_{\rm c}$ is a cut off in the momentum space.)
	Finally, the total kernel becomes
\begin{align}
	Q_{xx} (0)&=\frac{e^{2}N}{m}\pi T\sum_{n}
	\left[
	\frac{1}{|\ve_{n}|^{3}}-6\frac{\mD_0}{|\ve_{n}|^{5}}
	\right]
	|\Delta_\mu (\ve_{n})|^{2}.
	\label{sin_condition}
\end{align}
	From this result, it is revealed that the Meissner kernel is reduced by the critical spin fluctuation both for the even- and \oddw pairing.
	When the spin fluctuation is moderate, $\mD_0 \lesssim (\pi \Tc)^2/6$ (i.e., $g\lesssim \pi \sqrt{\Tc T_{\rm N}/6}$), the Meissner kernel becomes small, but remains positive.
	On the other hand, there is a possibility that the Meissner kernel becomes negative when the critical spin fluctuation is so strong, i.e., $\mD_0 \gtrsim (\pi \Tc)^2/6$.
	However, this rapid reduction is relaxed in we take into account the higher order corrections as is shown below in \S5.

\subsection{Dependences on band structure}
	In the derivation of eq. (\ref{15}), we assume that the density of state (DOS) is constant with respect to $\xi_{\bp}$.
	However, the DOS somewhat depends on $\xi_{\bp}$ in general.
	Here, we see the validity of the above result by expanding the DOS up to $\mathcal{O}(\xi^2)$ as
\begin{align}
	\sum_{\bp}v_p^2 F(\xi) =
	\frac{2}{3m}\int\!\! d\xi \NF \eF (1+ a_1 \xi +a_2 \xi^2 + \cdots) F(\xi).
\end{align}
	The $\xi$-linear term in DOS gives no contribution since $F(\xi)$ is an even function in $\xi$ for the relevant quantity, eq. (\ref{13}).
	For the zeroth order of $\mD_0$ term (eqs. (\ref{7})), the contribution from the $\xi^2$-term gives
\begin{align}
	\sum_{\bp}& v_p^2\left.\left(-2G_p^3 G_{-p} + G_p^2 G_{-p}^2 \right)
	\right|_{\mathcal{O}(\xi^2)}
	\nonumber\\
	&=
	\left[-2\left( \frac{a_2 \pi}{2} \right) + a_2 \pi \right] \frac{1}{|\ve_n|}
	=0,
\end{align}
	and for the $\mathcal{O}(\mD_0^1 )$ term, the contribution from the $\xi^2$-term becomes
\begin{align}
	&\sum_{\bp} v_p^2\left.\left(-12G_p^5G_{-p}^1 + 12 G_p^4 G_{-p}^2-6G_p^3G_{-p}^3 \right)
	\right|_{\mathcal{O}(\xi^2)}
	\nonumber\\
	&=
	\left[-12\left( -\frac{a_2 \pi}{16}\right) + 12 \times 0 -6\frac{a_2\pi}{8} \right] \frac{1}{|\ve_n|^3}=0.
\end{align}
	Consequently, the correction to $Q_{xx}$ is exactly zero up to $\xi^2$, i.e., the present result does not depend on the details of the band structure.

\section{Triplet pairings}
	Next, let us consider the case of triplet pairing.
	The triplet pairing is mediated by the ferromagnetic spin fluctuation for both the even- and odd-frequency.
	(The detailed discussions for the relationship between the pairing symmetry and the spin fluctuation are given in Appendix\ref{AppendixB}.)
	%
	We consider the ferromagnetic spin fluctuation in the form
\begin{align}
	\mD(\bq, \omega)=\frac{g^2 N_{\rm F}\kappa_0^2}{\kappa^2 + q^2  - \im \omega /\eta (q)},
\end{align}
	where $\kappa$ and $\kappa_0$ are the inverse correlation lengths with and without magnetic correlations, respectively.
	$\eta (q)$ is defined as $\eta (q)=T_{\rm sf}q$ using a characteristic spin-fluctuation temperature $T_{\rm sf}$\cite{Monthoux1999}.

	With the same procedure as for the singlet pairing, we obtain the correction to the kernel due to the ferromagnetic transverse spin-fluctuation as follows:
\begin{align}
	&\Delta Q_{xx}(q)=
	2e^{2}T\sum_{p, q'} \mD(q')
	\nonumber\\
	&\times
	\biggl[
	-v_{p}v_{-p}
	\mG(p_{+})\mG(p_{+}-q')\mF_{\up \up}^{+}(p_{-})\mF_{\up \up}(p_{+})
	\tag{a-1}
	\nonumber\\
	&-v_{p}v_{-p}
	\mG(-p_{+})\mG(-p_{+}+q')\mF_{\up\up}^{+}(p_{-})\mF_{\up\up}(p_{+})
	\tag{a-2}
	\nonumber\\
	&-v_{p}v_{-p}
	\mG(p_{-})\mG(p_{-}+q')\mF_{\up\up}^{+}(p_{-})\mF_{\up\up}(p_{+})
	\tag{a-3}
	\nonumber\\
	&-v_{p}v_{-p}
	\mG(-p_{-})\mG(-p_{-}+q')\mF_{\up\up}^{+}(p_{-})\mF_{\up \up}(p_{+})
	\tag{a-4}
	\nonumber\\
	&-v_{p}v_{-p+q'}
	\mG(p_{+})\mG(p_{-})\mF^{+}_{\down \down}(p_{+}-q')\mF_{\down \down}(p_{-}-q')
	\tag{d-1}
	\nonumber\\
	&-v_{p}v_{-p+q'}
	\mG(-p_{+}+q')\mG(-p_{-}+q')\mF_{\up\up}^{+}(p_{-})\mF_{\up\up}(p_{+})
	\tag{d-2}
	\nonumber\\
	&+v_{p_{1}}v_{p_{1}}
	\mG(p_{1+})\mG(p_{1+})\mG(p_{1-})\mG(p_{1+}-q')
	\tag{e-1}
	\nonumber\\
	&+v_{p}v_{p}
	\mG(p_{+})\mG(p_{-})\mG(p_{-})\mG(p_{-}+q')
	\tag{e-2}
	\nonumber\\
	&-v_{p}v_{p}
	\mG(p_{-})\mG(-p_{+}+q')\mF_{\up\up}^{+}(p_{+})\mF_{\up\up}(p_{+})
	\tag{f-1}
	\nonumber\\
	&-v_{p}v_{p}
	\mG(p_{+})\mG(-p_{-}+q')\mF_{\up\up}^{+}(p_{-})\mF_{\up\up}(p_{-})
	\tag{f-2}
	\nonumber\\
	&+v_{p}v_{p-q'}
	\mG(p_{+})\mG(p_{-})\mG(p_{+}-q')\mG(p_{-}-q')
	\tag{h-1}
	\nonumber\\
	&-v_{p}v_{-p}
	\mF_{\up\up}^{+}(p_{-})\mF_{\up\up}(p_{+})\mF_{\down\down}^{+}(p_{+}-q')\mF_{\up\up}(p_{+})
	\tag{k-1}
	\nonumber\\
	&-v_{p}v_{-p}
	\mF_{\up\down}^{+}(p_{-})\mF_{\down\up}(p_{+})\mF_{\up\down}^{+}(p_{-})\mF_{\down\up}(p_{-}-q')
	\tag{k-2}
\end{align}
	The corresponding diagrams are the same as that of singlet pairing except for (b), (c), (g), and (i) type diagrams in Fig. \ref{GFdiagrams}.
	(Of course, the spin indices are different from the singlet case.) 
	Up to the second order of $\Delta(p)$,
\begin{align}
	&\Delta Q_{xx}(0)
	=
	2e^{2}T\sum_{p}v_{p}^{2}\mD_0 |\Delta_\mu (p)|^{2}
	\nonumber\\ &\times 
	\left[
	-12G(p)^{5}G(-p)^{1}
	+6G(p)^{4}G(-p)^{2}
	-2G(p)^{3}G(-p)^{3}
	\right]
	\nonumber\\
	&\simeq 
	2e^{2}T\sum_{n}
	\mD_0 |\Delta_\mu (\ve_{n})|^{2}
	\frac{2N_{F}\ve_{F}}{3m}
	\nonumber\\&\times
	\left\{
	-12\left(\frac{\pi}{16}\right)
	+6\left(-\frac{\pi}{4}\right)
	-2\left(\frac{3\pi}{8}\right)
	\right\} \frac{1}{|\ve_{n}|^{5}}
	\nonumber\\&
	\simeq\frac{e^{2}N}{m}T\sum_{n}
	\mD_0 |\Delta_\mu (\ve_{n})|^{2}
	\left( -3\pi \right)
	\frac{1}{|\ve_{n}|^{5}}.
\end{align}
	Here, $\mD_0\sim g^2 N_{\rm F}\Tc \kappa_0^2 / \kappa^2$.
	The total kernel for triplet pairing is then obtained as
\begin{align}
	Q_{xx}(0)&=\frac{e^{2}N}{m}\pi T\sum_{n}
	\left[
	\frac{1}{|\ve_{n}|^{3}}-3\frac{\mD_0}{|\ve_{n}|^{5}}
	\right]
	|\Delta_\mu (\ve_{n})|^{2}.
	\label{tri_condition}
\end{align}
	The Meissner kernel for triplet pairing is also reduced by the spin fluctuation both for even- and \oddw, although the coefficient of the second term of eq. (\ref{tri_condition}) is smaller than that for singlet pairing.

\section{Higher order corrections}
	So far, we have shown that the coefficient of $\Delta Q_{xx}$ is reduced at least up to the first order of $\mD$.
	However, it is naturally expected that the signs of the higher order corrections are oscillating; when the sign of the first order is negative, that of the second order will be positive, and that of the third order will be negative etc..
	Thus, we have to check the effect of higher order corrections, especially paying attention to their convergence.
	Unfortunately, it is too complicated to proceed to the higher order corrections with the scheme we took in the previous sections.
	Here, we instead take a strategy where we calculate the superfluid density by a different method and then determine the Meissner kernel.	

	The superfluid density $\rho_{\rm s}$ is expressed in terms of the Meissner kernel as
\begin{align}
	\rho_{\rm s}(T) = \frac{N_{\rm s}(T)}{N}=\frac{m}{e^2 N}Q_{xx}.
\end{align}
	On the other hand, $\rho_{\rm s}$ can also be obtained from the spatially gradient term, $|\nabla \Delta (\br)|^2$, in the free energy\cite{Leggett1973,Leggett1998}, in other words, the coefficient of the $q^2$-term.
	This spatially gradient term is expressed by the superconducting pair susceptibility, $K_\text{p-p}(q)$, in the form
\begin{align}
	\rho_{\rm s}(T) = \frac{4m}{N}\sum_p |\Delta_\mu (p)|^2\frac{\partial ^2}{\partial q^2}K_\text{p-p}(q) .
\end{align}
	%
	%
	%
	We should have the same results for the Meissner kernel through the calculation of $K_\text{p-p}(q)$.
	In this section, we first show that the results in the previous sections can also be derived from the calculation of $K_\text{p-p}(q)$.
	Then, we see the effect of higher order corrections to the Meissner effect.
	%

\subsection{Up to 1st order}
	
	The spatial derivative term without the spin fluctuation is
\begin{align}
	\frac{\partial^2}{\partial q^2} K_\text{p-p}(q)&=
	-T\sum_{p}
	\frac{\partial^2}{\partial q^2}
	G_{p_+}G_{-p_-}
	\nonumber\\&
	=T\sum_p
	\biggl[
	\frac{v_p^2}{2} G(p)^2G(-p)^2
	-v_p^2G(p)^3 G(-p)
	\biggr]
	\nonumber\\
	&=
	\frac{N}{4m} \pi T \sum_n \frac{1}{|\ve_n |^3} .
\end{align}
	The corrections due to the spin fluctuation, which is the same one as in the previous sections, are given by the Feynman diagrams in Fig. \ref{higher} (a) and (b).
	Their contributions are 
\begin{align}
	\Delta K_\text{p-p}''(q) &= T\sum_p \mD_0 \frac{v_p^2}{2}
	\biggl[ \left( 6G_p^4 G_{-p}^2 -4 G_p^3 G_{-p}^3 \right)
	\nonumber\\&
	-2\left(
	6G_p^5 G_{-p} -3 G_p^4 G_{-p}^2 +G_p^3 G_{-p}^3
	\right)\biggr]
	\nonumber\\
	&\simeq
	-\frac{N}{4m}\pi T\sum_n \frac{6\mD_0}{|\ve_n|^5}
\end{align}
	for singlet pairings, and
\begin{align}
	\Delta K_\text{p-p}''(q) &= T\sum_p \mD_p \frac{v_p^2}{2}
	\bigl[ 
	-12G_p^5 G_{-p} +6 G_p^4 G_{-p}^2 
	-2G_p^3 G_{-p}^3
	\bigr]
	\nonumber\\
	&\simeq
	-\frac{N}{4m}\pi T\sum_n \frac{3\mD_0}{|\ve_n|^5}
\end{align}
	for triplet pairings.
	Then the total $\rho_{\rm s}(T)$ up to $\mathcal{O}(\mD^1)$ are
\begin{align}
	\rho_{\rm s}^{\rm sin} (T)
	&\simeq \pi T\sum_n
	\frac{1}{|\ve_n|^3}
	\left[ 1-\frac{6\mD_0 }{|\ve|^2}\right] |\Delta_\mu (\ve_n)|^2,
	\\
	\rho_{\rm s}^{\rm tri}(T)
	&\simeq \pi T\sum_n
	\frac{1}{|\ve_n|^3}
	\left[ 1-\frac{3\mD_0 }{|\ve|^2}\right] |\Delta_\mu (\ve_n)|^2,
\end{align}  
	for singlet and triplet pairings, respectively.
	These results are exactly equivalent to the results obtained from the current-current response function in the previous sections, eqs. (\ref{sin_condition}) and (\ref{tri_condition}).
	%

\subsection{2nd and 3rd order}
\begin{figure}[tb]
\begin{center}
\includegraphics[width=8cm]{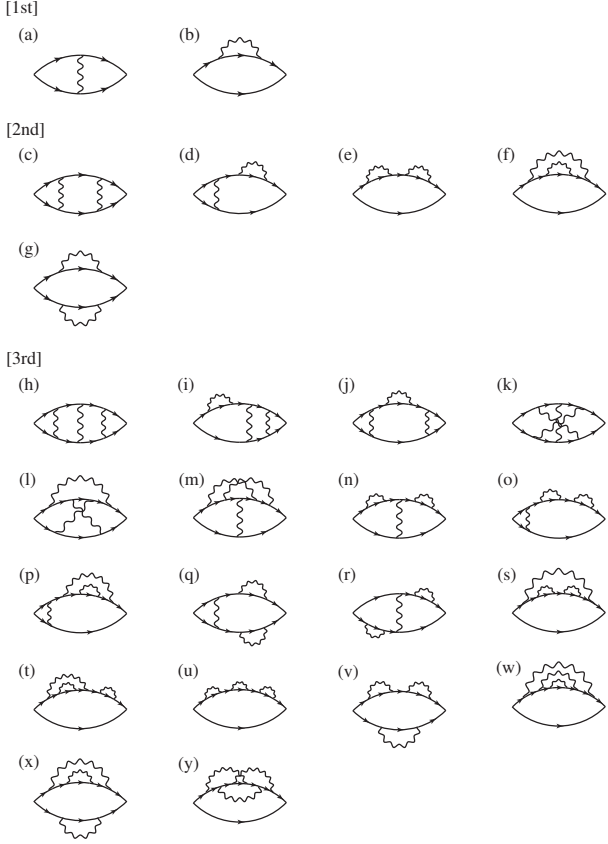}
\end{center}
\caption{Diagrammatic expressions of the pair susceptibility with correction of fluctuations.}
\label{higher}
\end{figure}
	%
	Now, we have confirmed the equivalence between two different procedures for calculating $\rho_{\rm s}(T)$ up to the first order of spin fluctuation, $\mathcal{O}(\mD^1)$.
	With regard to the higher order calculations, we can carry out the calculations of the spatially gradient term more easily than that of the current-current response functions.
	%
	%
	The results for singlet pairing are as follows:
\begin{align}
	\Delta^{(2)} K_\text{p-p}''(q)&=-
	T\sum_{p,q} \mD_q^2
	\nonumber\\
	&\times
	\frac{\partial^2}{\partial q^2}
	\biggl[
	G_{p_+}^3 G_{-p_-}^3
	-4G_{p_+}^4 G_{-p_-}^2
	+2G_{p_+}^5 G_{-p_-}^1
	\nonumber\\&
	+2G_{p_+}^5 G_{-p_-}^1
	+G_{p_+}^3 G_{-p_-}^3
	\biggr]
	\nonumber\\
	&=
	\frac{N}{4m}\pi T\sum_n
	\frac{30 \mD_0^2}{|\ve_n|^7},
\end{align}
	for the second order (Fig. \ref{higher}(c)-(g)), and 
\begin{align}
	\Delta^{(3)} K_\text{p-p}''(q)&=-
	T\sum_{p,q} \mD_q^2
	\nonumber\\
	&\times
	\frac{\partial^2}{\partial q^2}
	\biggl[
	-G_{p_+}^4 G_{-p_-}^4
	+4G_{p_+}^5 G_{-p_-}^3
	+2G_{p_+}^5 G_{-p_-}^3
	\nonumber\\&
	-2G_{p_+}^4 G_{-p_-}^4
	+2G_{p_+}^5 G_{-p_-}^3
	-2G_{p_+}^6 G_{-p_-}^2
	\nonumber\\&
	-2G_{p_+}^6 G_{-p_-}^2
	-4G_{p_+}^6 G_{-p_-}^2
	-4G_{p_+}^6 G_{-p_-}^2
	\nonumber\\&
	-2G_{p_+}^4 G_{-p_-}^4
	-2G_{p_+}^4 G_{-p_-}^4
	+2G_{p_+}^7 G_{-p_-}^1
	\nonumber\\&
	+4G_{p_+}^7 G_{-p_-}^1
	+2G_{p_+}^7 G_{-p_-}^1
	+2G_{p_+}^5 G_{-p_-}^3
	\nonumber\\&
	+2G_{p_+}^7 G_{-p_-}^1
	+2G_{p_+}^5 G_{-p_-}^3
	+2G_{p_+}^7 G_{-p_-}^1
	\biggr]
	\nonumber\\
	&=
	-\frac{N}{4m}\pi T\sum_n
	\frac{168 \mD_0^3}{|\ve_n|^9}, 
\end{align}
	for the third order (Fig. \ref{higher} (h)-(y)).
	The total singlet pair susceptibility up to third order is given by
\begin{align}
	K_\text{p-p}''(0)=
	\frac{N}{4m} \pi T&\sum_n 
	\frac{1}{|\ve_n|^3}
	\Biggl[
	1-6\frac{\mD_0 }{|\ve_n|^2}
	+30\left( \frac{\mD_0 }{|\ve_n|^2}\right)^2
	\nonumber\\&
	-150 \left( \frac{\mD_0 }{|\ve_n|^2}\right)^3
	-18 \left( \frac{\mD_0 }{|\ve_n|^2}\right)^3
	\Biggr].
\end{align}
	%
	%
	Rather surprisingly, the coefficients of each term are integers (although each term is given as a fractional number, e.g., $G_p^5G_{-p}^5=35\pi/128|\ve_n|^9$, or $G_p^4 G_{-p}^6=-7\pi/32|\ve_n|^9$).
	Moreover, they are almost given as a geometric progression, which can be written in the convergent form:
\begin{align}
	\rho_{\rm s}&\simeq
	\pi T\sum_n 
	\frac{1}{|\ve_n|^3}
	\left[
	1-\frac{6 (\mD_0 /|\ve_n|^2)}{1+5(\mD_0 /|\ve_n|^2)}
	\right]
	|\Delta_\mu (\ve_n )|^2
	\label{32}
\end{align}
	%
	%
	Therefore, $\rho_{\rm s}$ and so the Meissner kernel is reduced by the spin fluctuations, even if we take into account the higher order corrections.
	With this geometric progression form, we can go beyond the perturbation theory even though we made calculations only up to the third order.
	The results should be valid as far as it converges, so that our results (\ref{32}) is expected to be valid for any magnitude of the spin fluctuations, since it is written in a convergent form.
	Then, we also conclude that the Meissner kernel can be negative both for even- and \oddw when the effect of massless spin fluctuations is sufficiently large.
		
	Similar estimations of the higher order corrections will be given for the triplet pairing just by changing the coefficients.
	The result so obtained will be qualitatively the same as that for the singlet pairing, namely, the Meissner kernel of the \oddw triplet pairing also becomes positive with the critical spin fluctuations.

\section{Discussions}
\begin{figure}[tb]
\begin{center}
\includegraphics[width=7cm]{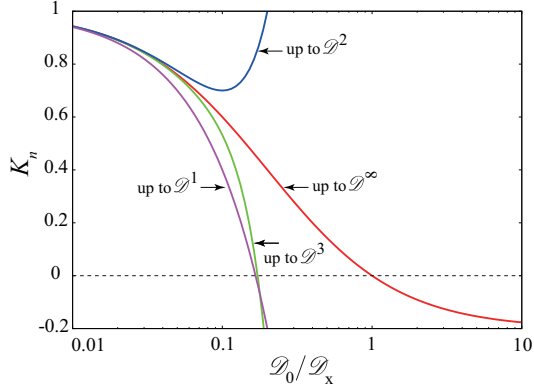}
\end{center}
\caption{(Color online) Relation between the spin fluctuation $\mD_0$ and the Meissner kernel, which corresponds to the super fluid density $\rho_{\rm s}$ or the square of the inverse penetration depth $\lambda_{\rm L}^{-2}$.
}
\label{illust}
\end{figure}

	Summarizing the progress hitherto made, we find that the Meissner kernel in the GL region is given by
\begin{align}
	Q_{xx}&=\frac{e^2 N}{m} \frac{\pi T}{|\ve_n|^3} \sum_n
	\left[
	1- \frac{6\mD_0 / |\ve_n|^2}{1+5\mD_0 /|\ve_n|^2}
	\right] |\Delta_\mu (\ve_n)|^2
	, 
\end{align}
	whose $\mD_0$ dependence is depicted in Fig. \ref{illust}.
	%
	%
	%
	This conclusion will be valid, even if we consider the infinite order in $\mD_0$.
	The Meissner kernel is positive but strongly reduced by the critical spin fluctuations both for even- and \oddw pairing.
	Meanwhile, our results also indicate that the Meissner kernel can become negative both for even- and \oddw pairing, when the critical fluctuations are sufficiently large. 
	In eq. (\ref{32}), the dominant contribution from the $\ve_n$-summation comes only from the lower $\ve_n$, since $\Delta (\ve_n)$ is rapidly suppressed as $\ve_n$ increases when the retardation of the interaction is strong\cite{Fuseya2003,Kusunose2010b}.
	Then, the condition for the sign change is roughly given by $ \mD_0 \gtrsim \Tc^2 ( \equiv \mD_{\rm x})$ or $g\gtrsim \sqrt{\Tc T_{\rm N}}$ in the case of singlet pairing in the AF phase.
	Therefore, we can conclude that with moderate magnitude of spin fluctuations, $\mD_0 < \mD_{\rm x}$, the superconductivity with zero center-of-mass momentum is stable, but the superfluid density is strongly reduced, while with sufficiently large spin fluctuations, $\mD_0 > \mD_{\rm x}$, the uniform superconductivity is no longer stable.
	In such a situation of $\mD_0 > \mD_{\rm x}$, the superconductivity with Cooper pairs of the finite center-of-mass momentum can be stabilized, since the coefficient of $q^2$-term is negative, but that of $q^4$-term is positive.
	Then, the FFLO pairing can be realized spontaneously {\it without the magnetic field}.

	In the following discussion, we make some comments on the relationship between the present results and the previous works.
	The qualitative tendency of the present results are consistent with the standard theory of a superfluid Fermi liquid\cite{Leggett1965}, where $\lambda_{\rm L}$ is given by
\begin{align}
	\lambda_{\rm L}^{-2}(T)&=
	\frac{4\pi Ne^2}{m^*c^2} {\textstyle (1+\frac{1}{3}F_1^{\rm s})}\frac{1-Y(T)}{1+\frac{1}{3}F_1^{\rm s} Y(T)},
	\label{Leggett}
\end{align}
	where $m^*$ is the effective mass of quasiparticles, $F_1^{\rm s}$ is the Landau parameter, and $Y(T)$ is the so-called Yosida function. 
	%
	%
	It is noted that eq. (\ref{Leggett}) is also valid for the systems without the Galilean invariance.
	There, the incoherent contribution for the current-current response function is not negligible and is incorporated into the Fermi liquid parameter $F_1^{\rm s}$, the $\omega$-limit vertex $\Gamma^{\omega}$, which arises from incoherent processes\cite{Leggett1965}.
	In the case of correlated electron systems without the translational invariance, such as the heavy fermion systems, eq. (\ref{Leggett}) is modified as the effective mass $m^*$ is replaced by the dynamic effective mass $m_{\rm d}$\cite{Varma1986}.
	($m_{\rm d}$ is given as $m_{\rm d}/m = \left[ 1-\partial \Sigma (k, \omega) /\partial \omega \right]_{k=k_{\rm F}, \omega=E_{\rm F}}$, where $\Sigma (k, \omega)$ is the self energy.)
	The spin-fluctuation increases $m_{\rm d}$, so that it reduces $\lambda_{\rm L}^{-2}$.
	When the system is near the AF critical point or the metal-insulator transition, the Drude weight, $D=(\pi e^2 N/m^*) (1+\frac{1}{3}F_1^{\rm s})$, is reduced\cite{Baeriswyl1987,Okabe1998,Fuseya2000,Miyake2001}.
	In this case, $\lambda_{\rm L}^{-2}$ is also reduced, which is consistent with our results.
	However, in the derivation of eq. (\ref{Leggett}), the possibility of the FFLO state is discarded, so that we cannot discuss the possibility of FFLO state only from eq. (\ref{Leggett}).
	On the other hand, in our theory, we directly calculate the coefficient of $q^2$-term, which corresponds to the order parameter of the FFLO state, so that we can discuss the possibility of the FFLO state, resulting in the possible zero-field FFLO state.

	Abrahams {\it et al.} (and more recently Dahal {\it et al.}) introduced a composite operator which couples a Cooper pair to a spin fluctuation and has the same symmetry as the \oddw pairing\cite{Abrahams1995,Dahal2009}.
	They showed that, in the composite-operator condensation states, the superfluid density is reduced in some models of the magnon propagators.
	Although the magnon propagator in Ref. \citen{Abrahams1995}, $\mathcal{D}(q)$, resembles our spin fluctuation, $\mD(q)$, their contributions to the Meissner kernel are different from ours.
	For example, in the first order of $\mathcal{D}(q)$ or $\mD (q)$, the composite operator theory leads to $\mathcal{D}(q)\left[-2G_p^3 G_{-p} +G_p^2 G_{-p}^2\right]$ (eq. (4.3) in Ref. \citen{Abrahams1995}), which has the form of the first line of eq. (\ref{7}) supplemented by the magnon propagator, whereas $\mathcal{O}(\mD^1)$-term is given by $\mD(q')\left[-12 G_p^5 G_{-p}^1+12G_p^4 G_{-p}^2 -6G_p^3 G_{-p}^3\right]$ (the first line of eq. (\ref{13})) in our theory. 
	Furthermore, in our theory, the higher order terms are shown to converge having an almost geometric progression structure, even though the sign of the higher order of $\mD$ is fluctuating.
	However, in the composite operator theory, the contribution of the higher order term has not been investigated as to whether it converges or diverges; it can differ greatly from the present results.
	Nevertheless, both approaches share a common aspect in which incoherent degrees of freedom beyond the quasiparticle picture (in a sense of the Fermi liquid theory) play crucial roles.

	Finally, we comment on the \oddw in the non-uniform system, such as the normal metal / superconductor junctions.
	For the Meissner effect of the pairs in the normal metal region, we do not need to consider the corrections due to the incoherent part, since the spin fluctuations are irrelevant there.
	%

\section{Conclusions}
	
	The effects of the critical spin fluctuations, the crucial ingredient for realizing the \oddw superconductivity, on the electromagnetic response of the \oddw superconductivity have been examined.
	The Meissner kernel is strongly reduced by the critical spin fluctuations both for the even- and \oddw pairing.
	Thus the superfluid density will be smaller and the penetration depth will be longer than the case without corrections due to the critical spin fluctuations.
	Moreover, when the effect is strong enough, the Meissner kernel becomes negative, suggesting the zero-field spontaneous FFLO state.
	This conclusion has been reached by calculating two different values: one is the current-current response function,\cite{AGD} and the other is the spatially gradient term (the coefficient of $q^2$-term) in the free energy.\cite{Leggett1973,Leggett1998}
	The results, obtained by two different methods, completely agree with each other at least up to the first order of fluctuation.
	This is valid both for singlet pairing with the antiferromagnetic fluctuation in the coexisting phase and for triplet pairing with the ferromagnetic fluctuation.
	The validity has been also checked for the shape of the density of state in the case of singlet pairing, and the present results are shown to be invariant, namely, the present conclusion is not affected by the detail of the band structure.
	%

\section*{Acknowledgments}
	
	We are grateful to H. Kusunose for fruitful discussions.
	This work is supported by a Grant-in-Aid for Scientific Research (No. 19340099) from The Japan Society for the Promotion of Science (JSPS), and a Grant-in-Aid for Scientific Research on Innovative Areas ``Heavy Electrons" (No. 20102008) from the Ministry of Education, Culture, Sports, Science and Technology, Japan.
	One of the authors (Y. F.) is supported by a Grant-in-Aid for Young Scientists（Start-up） from JSPS.

\appendix

\appendix
\section{Possible symmetries of \oddw pairing}
\label{AppendixB}
\begin{figure}[tbp]
	\begin{center}\leavevmode
	\rotatebox{0}{\includegraphics[width=80mm]{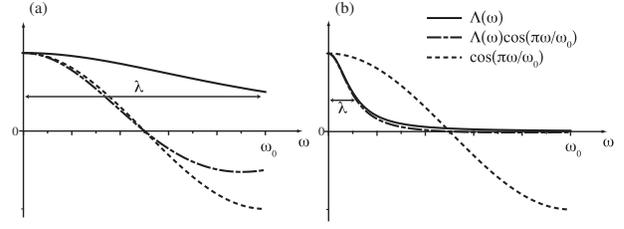}}
	\caption{Illustration of the $\omega$-dependent pairing interaction $\Lambda (\omega)$ and 
	the function $\Lambda (\omega )\cos (\pi \omega /\omega_0 )$ for (a) $\eta \sim \omega_0$ 
	and (b) $\eta \ll \omega_0$. The coefficient $\lambda_0$ is given by the integration of $\Lambda (\omega )$
	and $\lambda_1$ by that of $\Lambda (\omega )\cos (\pi \omega /\omega_0 )$.
	We can see from these illustrations that $\lambda_0 \gg \lambda_1$ for the instantaneous interaction
	and $\lambda_0 \sim \lambda_1$ for the retarded interaction.
}\label{Vwcosab}\end{center}\end{figure}

	Here, we discuss the possible symmetries of the \oddw pairing on the basis of a general argument.
	It is found that the \oddw pairing becomes competitive with the \evenw one
	when the interaction is strongly retarded.
	We assume that the pairing interaction $V({\bf q}, \omega )$ can be separated as
	\begin{eqnarray}
	V({\bf q}, \omega )=\Phi ({\bf q}) \Lambda (\omega ),
	\end{eqnarray}
	where the dimensionality of the system is arbitrary.
	\begin{figure}[tbp]
\begin{center}\leavevmode
\rotatebox{0}{\includegraphics[width=60mm]{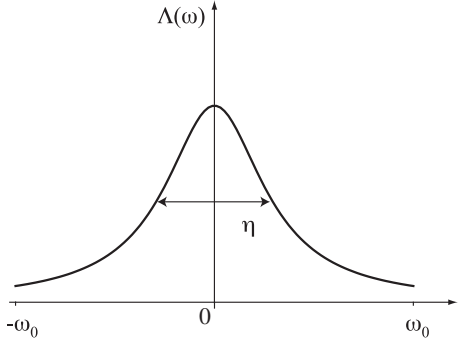}}
\caption{Illustration of the frequency dependence of pairing interaction $\Lambda (\omega)$.
}\label{Vwcos}\end{center}\end{figure}
	Generally, we can assume that $\Lambda (\omega )$ has a maximum at $\omega=0$,
	and we parameterize its peak width by $\eta$ as displayed in Fig. \ref{Vwcos}.
	Since $\Lambda (\omega )$ is an even function,
	it can be expanded as
	\begin{eqnarray}
	\Lambda (\omega )&=&\frac{\lambda_0}{2}+\lambda_1 \cos \frac{\pi \omega }{\omega_0}
	+\lambda_2 \cos \frac{2\pi \omega}{\omega_0} +\cdots , \\
	\lambda_n &=& \frac{2}{\omega_0}\int_0^{\omega_0} \!\! d\omega \, 
	\Lambda (\omega )\cos \frac{n\pi \omega }{\omega_0},
	\end{eqnarray}
	with a sufficiently large cut-off $\omega_0$.
	The pairing interaction is written in the separable form
	\begin{eqnarray}
	\Lambda^{\rm even}(\varepsilon -\varepsilon ')&=& \frac{\lambda_0}{2}+
	\lambda_1 \cos \frac{\pi \varepsilon}{\omega_0}\cos \frac{\pi \varepsilon '}{\omega_0} +\cdots , \\
	\Lambda^{\rm odd}(\varepsilon -\varepsilon ')&=& 
	\lambda_1 \sin \frac{\pi \varepsilon}{\omega_0}\sin \frac{\pi \varepsilon '}{\omega_0} +\cdots ,
	\end{eqnarray}
	for the pair scattering 
	$(\varepsilon , -\varepsilon ) \to (\varepsilon ', -\varepsilon ')$.
	The maximum eigenvalue
	is approximately given by $\lambda_0 /2$ for even-frequency pairing 
	and $\lambda_1$ for odd-frequency pairing.
	If $\Lambda (\omega)$ is independent of $\omega$, that is to say, $\eta \to \infty$, $\lambda_0 >0$ 
	and $\lambda_n =0$ for all $n\ne 0$.
	When $\Lambda (\omega)$ is a slowly varying function, i.e., $\eta\sim \omega_0$, $\lambda_1$ becomes finite but very small as is seen in Fig. \ref{Vwcosab} (a).
%
%
	Therefore, such instantaneous interaction only mediates the \evenw pairing.
	On the other hand, in cases where $\Lambda (\omega )$ strongly depends on $\omega$, i.e., $\eta \ll \omega_0$,
	the function $\Lambda (\omega )\cos (\pi \omega /\omega_0 )$ has almost the same form
	as $\Lambda (\omega )$, so that $\lambda_1$ is nearly equal to $\lambda_0$.
	Then, $\Tc$ of the \oddw pairing
	can be comparable to
	that of the \evenw pairing.
	In fact, if the momentum part of \oddw $\phi ^{\rm odd}({\bf q})$, which is neglected so far,
	has a larger value than that of \evenw $\phi ^{\rm even} ({\bf q})$,
	it is possible that the \oddw pairing is stabilized prevailing over the \evenw pairing.
	In order to determine which $\Tc$ is higher and what type of gap structure develops
	below $\Tc$, we need more data about the model.

\subsection{Spin fluctuation model}

\begin{figure}[tbp]
	\begin{center}\leavevmode
	\rotatebox{0}{\includegraphics[width=80mm]{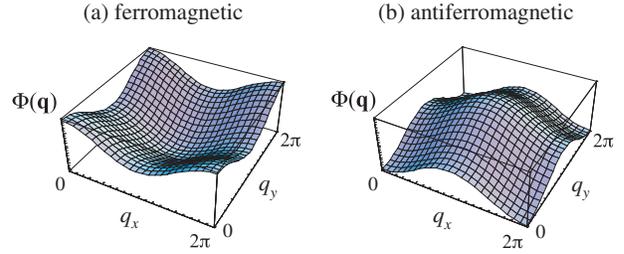}}
	\caption{Illustration of the momentum dependence of the pairing interaction $\Phi ({\bf q})$
	for (a) ferromagnetic cases and for (b) antiferromagnetic cases.
}\label{modelVq}\end{center}\end{figure}

	Now, we consider the varieties of pairing
	on the basis of the two-dimensional spin fluctuation model.
	Let us use 
	\begin{align}
	V_{\rm s}({\bf q}, \omega ) &= 3\Phi ({\bf q})\Lambda (\omega ), \\
	V_{\rm t}({\bf q}, \omega ) &= -\Phi ({\bf q})\Lambda (\omega ),
	\end{align}
	as the model interaction for singlet $(V_{\rm s})$ and triplet $(V_{\rm t})$ pairing
	with the momentum dependence of $\Phi ({\bf q})$ given as
	\begin{align}
	\Phi ({\bf k-k'}) = \phi_0 + 2\phi_1 [\cos (k_x -k_x' ) +\cos (k_y -k_y' )],
	\end{align}
	where the ferromagnetic spin fluctuations gives $\phi_0 > \phi_1 >0$
	and the antiferromagnetic spin fluctuation $\phi_0 > -\phi_1 >0$. (See Fig. \ref{modelVq})
%
%
	
	We can express $\Phi ({\bf k-k'})$ in the separable form\cite{Miyake1986}
	\begin{align}
	\Phi^{\rm even} ({\bf k-k'}) &= \phi_0 +2\phi_1 (\gamma_k \gamma_{k'} +\eta_k \eta_{k'} ), \\
	\Phi^{\rm odd} ({\bf k-k'}) &= 2\phi_1 (\sin k_x \sin k_x '+\sin k_y \sin k_y '),
	\end{align}
	where $\gamma_k = (\cos k_x + \cos k_y)/\sqrt{2}$ 
	and $\eta_k = (\cos k_x -\cos k_y )/\sqrt{2}$.
	The functions $\gamma_k$, $\eta_k$, and $\sin k_{x, y}$ are irreducible representations
	of the symmetry group of the square lattice.
	Then, adding the frequency part 
	$\Lambda (\omega ) = \lambda_0 /2 + \sum_n \lambda_n \cos (n\pi \omega / \omega_0 )$,
	the pairing interactions have the forms
	\begin{align}
	V^{\rm e, e}_{\rm s}(k-k') 
	&= 3\left[  \phi_0 +2\phi_1 (\gamma_k \gamma_{k'} +\eta_k \eta_{k'} )\right]
	\nonumber \\ &\times
	( \lambda_0 /2 + \lambda_1 \cos \epsilon \cos \epsilon ') , \\
	V^{\rm o, o}_{\rm s}(k-k') 
	&= 3\left[  2\phi_1 (\sin k_x \sin k_x '+\sin k_y \sin k_y ')\right]
	\nonumber \\ &\times
	(\lambda_1 \sin \epsilon \sin \epsilon ') , \\%
	V^{\rm e, o}_{\rm t}(k-k' ) 
	&= -\left[  \phi_0 +2\phi_1 (\gamma_k \gamma_{k'} +\eta_k \eta_{k'} )\right]
	\nonumber \\ &\times
	(\lambda_1 \sin \epsilon \sin \epsilon ') ,\\
	V^{\rm o, e}_{\rm t}(k-k' ) 
	&= -\left[  2\phi_1 (\sin k_x \sin k_x '+\sin k_y \sin k_y ')\right]
	\nonumber \\ &\times
	(\lambda_0 /2 + \lambda_1 \cos \epsilon \cos \epsilon ') , 
	\end{align}
	where $\epsilon = \pi \varepsilon /\omega_0 $, 
	and $V^{\rm e, e}_{\rm s}$ denotes the even-parity and \evenw singlet pairing interaction,
	$V^{\rm e, o}_{\rm t}$ denotes the even-parity and \oddw triplet pairing interaction, and so on.
%
\begin{table}
	\begin{center}
	\begin{tabular}{rccc}\hline
	{\bf ferro.}	& $s$-wave 	& $p$-wave 	& $d$-wave	\\ \hline
	singlet	& ---		& ---		& ---		\\ 
	triplet	& \framebox[10mm]{$\pi \phi_0 \lambda_1 $}	
	& $\phi_1 \lambda_0 $	& \framebox[10mm]{$2\phi_1 \lambda _1 $} \\ \hline
	\end{tabular}
	\end{center}
	\begin{center}
	\begin{tabular}{rccc}\hline
	{\bf antiferro.}	& $s$-wave 	& $p$-wave 	& $d$-wave	\\ \hline
	singlet	& ---	& \framebox[10mm]{$6\phi_1 \lambda_1 $}	& $3\phi_1 \lambda_0$	\\ 
	triplet	& \framebox[10mm]{$\pi \phi_0 \lambda_1$}	& ---	& --- \\ \hline
	\end{tabular}
	\end{center}
	\caption{Possible pairing symmetries and their coefficients of the attractive components 
	for ferromagnetic cases, $\phi_1 >0 $ (upper table),
	and antiferromagnetic cases, $\phi_1 <0 $ (lower table).
	The frame represents the \oddw gap.
	}
	\label{ferrotable}
	\end{table}
	
	In this representation, the negative coefficient corresponds to the attractive channel, so that only the irreducible representation with a negative coefficient can become the superconducting gap function.
	For the ferromagnetic case, i.e., $\phi_0 >\phi_1 >0$, both odd- and even-parity triplet pairings are possible, while the singlet pairing is not.
	The \evenw triplet pairing has the $p$-wave symmetry,
	and the \oddw triplet pairing has the $s$-wave symmetry or the $d_{x^2 -y^2} $-wave symmetry.
	The $\Tc$ for $p$-wave triplet is related to the eigenvalue $ \phi_1 \lambda_0 $
	and that for $s$-wave triplet and for $d_{x^2 -y^2}$-wave
	triplet pairing are related to $\pi \phi_0 \lambda_1$ and $2\phi_1 \lambda_1$,
	respectively.
	(See Table \ref{ferrotable} for ``ferro.".)
	Here, we omit the extend-$s$-wave pairing for the $\gamma$ components.
	Therefore, if $\lambda_1$ is comparable to $\lambda_0 /2$,
	\oddw $s$-wave triplet pairing is expected to arise since $\phi_0 > \phi_1$ in general.
	The \oddw $s$-wave triplet pairing that appears in the triangular lattice\cite{Vojta1999,Yada2008,Shigeta2009,Shigeta2010} would
	correspond to this case.

	Similarly, for the antiferromagnetic case, $-\phi_0 < \phi_1 <0$,
	the anisotropic even-parity singlet pairing, the odd-parity singlet pairing, 
	and the isotropic even-parity triplet pairing is possible.
	The $\Tc$ for the $d_{x^2 -y^2}$-wave singlet 
	is related to the eigenvalue
	$3\phi_1 \lambda_0$ and 
	that for the $p$-wave singlet is related to $6\phi_1 \lambda_1$.
	In the case of the $s$-wave triplet, the $\Tc$ is related to
	$\pi \phi_0 \lambda_1 $. (See Table \ref{ferrotable} for ``antiferro.".)
	Therefore, if $\lambda_1$ is comparable to $\lambda_0$,
	the $p$-wave singlet and the $d_{x^2 -y^2}$-wave singlet pairing are nearly degenerate.
	The $p$-wave singlet pairing suggested in Ref. \citen{Fuseya2003} would correspond to this case.
	
	To summarize, $s$-wave and $d$-wave triplet \oddw pairings are possible for the ferromagnetic case, and $s$-wave triplet and $p$-wave singlet \oddw pairings are possible for the antiferromagnetic case.


\end{document}